# Calorimetric measurements at low temperatures in toluene glass and crystal

C. Alvarez-Ney[1] • J. Labarga[1,2] • M. Moratalla[1] • J. M. Castilla[1] • M. A. Ramos[1,3]

[1] Laboratorio de Bajas Temperaturas, Departamento de Física de la Materia Condensada, Universidad Autónoma de Madrid, Madrid, Spain

[2] ETSI Caminos, Canales y Puertos, Universidad Politécnica de Madrid, Madrid, Spain

[3] Condensed Matter Physics Center (IFIMAC) and Instituto Nicolás Cabrera (INC), Universidad Autónoma de Madrid, Madrid, Spain

**Abstract** The specific heat of toluene in glass and crystal states, has been measured both at low temperatures down to 1.8 K (using the thermal relaxation method) and in a wide temperature range up to the liquid state (using a quasiadiabatic continuous method). Our measurements therefore extend earlier published data to much lower temperatures, thereby allowing to explore the low-temperature "glassy anomalies" in the case of toluene. Surprisingly, no indication of the existence of tunneling states is found, at least within the temperature range studied. At moderate temperatures, our data either for the glass or for the crystal show good agreement with those found in the literature. Also, we have been able to prepare bulk samples of toluene glass by only doping with 2% mol ethanol instead of with higher impurity doses used by other authors.

## 1 Introduction

Among the many unsolved puzzles in the field of glasses and the glass transition phenomenon [1], low-temperature properties and low-energy dynamics of glasses and non-crystalline solids in general, remain most controversial, in spite of much research performed in the last 45 years.

In 1971, Zeller and Pohl [2] demonstrated that low-temperature thermal properties of glasses did not follow at all the expected Debye behavior, as non-metallic crystals do. In brief, the specific heat of glasses depends approximately linear on temperature $C_p \propto T$ below 1 K, and the thermal conductivity increases quadratically $\kappa \propto T^2$, in clear contrast with the cubic dependences expected and observed in crystals, well understood in terms of Debye's theory. Besides, a broad peak in $C_p/T^3$ is ubiquitously observed around 5–10 K, above the expected Debye level, together with a universal plateau in the thermal conductivity that is orders of magnitude lower than that of their crystalline counterparts [2,3].

Very short after those seminal experiments by Zeller and Pohl, some theoreticians developed the now well-known Tunneling Model (TM) [4,5] to successfully explain these low-temperature anomalies, at least those below 1 K: the linear term in $C_p(T)$, the squared temperature dependence of $\kappa(T)$, and some others [3]. Essentially, the TM proposed the general existence of a random, constant distribution of asymmetric double-well potentials in amorphous solids due to its configurational disorder. These would correspond to the possibility that atoms or groups of atoms could move even at low temperatures by quantum mechanical tunneling from one configuration to another of similar potential energy.

But what occurs above 1 K? As Buchenau and co-workers first showed in the 80's combining inelastic neutron scattering with specific heat experiments [6,7], this maximum in $C_p/T^3$ is clearly originated by an excess of the vibrational density of states (VDOS) over the Debye lattice contribution derived from acoustic data. This so-called *boson peak*, that had been reported for decades in Raman-scattering at low frequencies in glasses [3,8,9], was then demonstrated just to be an ubiquitous broad peak in the Debye-reduced VDOS $g(\nu)/\nu^2$ and hence in $C_p/T^3$.

These thermal properties above 1 K and the vibrational dynamics of glasses around, say, 0.5–5 meV are much more controversial nowadays. Nevertheless, one of the most used models to account for the low-temperature properties of glasses also above 1 K is the Soft Potential Model (SPM), which postulates that the Debye theory is also applicable to glasses at very low temperature, only that they coexist with additional quasilocalized soft modes or low-energy excitations in non-crystalline solids [10,11]. The SPM is somehow an extension of the TM, where in addition of double-well potentials producing two-level systems (TLS), there are also more-or-less harmonic single-well potentials responsible for low-frequency vibrations (*soft modes*). These two zones are separated by an anharmonic quartic potential, which marks the crossover from the TLS-dominated lowest-temperature region to another one above a few K, dominated by those low-frequency vibrations with $g(\nu) \propto \nu^4$ –and hence $C_p \propto T^5$– [12]. Eventually, this rise of the VDOS over the Debye level saturates and the boson peak in

$g(\nu)/\nu^2$ arises [13], which marks the end of the range where acoustic phonons and quasilocalized soft modes can be considered as independent [14].

An alternative approach to account for the universal existence of the *boson peak* in the VDOS and the plateau in the thermal conductivity of glasses, is the theory of randomly fluctuating elastic constants [15], though it does not allow such a straightforward analysis of experimental data as with the SPM. We will therefore use a practical version of the SPM [12,16] to analyze our specific-heat measurements of toluene glass.

Notwithstanding the general acceptance of the standard TM mentioned above, Leggett and other authors [17,18] have raised strong criticisms against it, arguing how improbable is that a random ensemble of independent tunneling states or TLS would produce essentially the same universal constant for the thermal conductivity or the acoustic attenuation in any disordered solid. Nonetheless, most experimentalists have continued to use the TM, given its apparent success to account for most of the experimental data at very low temperatures. However, recent measurements of the specific heat in ultrastable glasses of indomethacin [19] have shown that the tunneling TLS unexpectedly dissappear, what was attributed to its very anisotropic and layered character. This experimental finding was claimed to lend support to the arguments against the standard TM mentioned above [17, 18].

Toluene is a simple and well-known organic substance. Its molecule is essentially a benzene ring with one methyl group, hence it is also named methylbenzene. The interest in measuring the low-temperature specific heat of glassy toluene is twofold: (i) to extend specific-heat data to lower temperatures in this much studied, very fragile (bad) glass-forming liquid; (ii) to check the suggestion by Leggett [20] that toluene glass could be a good benchmark of the TM, after the observation by Naimov et al. [21], using single-molecule spectroscopy, that the dynamics of several low-molecular-weight glasses as toluene did not follow the low-temperature behavior expected from TLS within the Tunneling Model.

In this paper we present our recent measurements of the specific heat of toluene, both in its crystal and glass states. In the temperature range roughly between 1.8 and 20 K, the standard thermal relaxation method was employed. In addition, a quasiadiabatic continuous method was used to cover the range up to 250 K in the liquid state.

## 2 Experimental

2.1 Materials and experimental techniques

Toluene (or methylbenzene, $C_6H_5CH_3$) was purchased from Sigma-Aldrich (purity: > 99.9%) and used without further purification. In order to be able to vitrify liquid toluene and avoid crystallization (see Fig. 1), some samples of toluene doped with 2% mol ethanol were also prepared, employing pure and dried ethanol (max. 0.02% $H_2O$) also without further purification.

The heat capacity of the samples was measured in a versatile calorimetric system developed in our laboratory [22] especially intended for glass-forming liquids. With this experimental system we are able to concurrently study and characterize the phase transitions in the range 77–300 K (first using liquid nitrogen as thermal sink), and then to measure their specific heat at low temperatures where the liquid nitrogen bath is readily replaced by liquid helium, eventually pumped to achieve temperatures about 1.5 K.

At temperatures below 30 K, the typical thermal relaxation method was used, whereas at higher temperatures a quasiadiabatic continuous method developed by us was employed [22,23]. Although the latter method has been preferentially used with liquid nitrogen as thermal bath, we showed more recently that it can also be employed at lower temperatures with liquid helium [23]. Many more details about the cryogenic system employed, electronic control, thermal sensors and heating elements can be found in [23]. In the present work, the calorimetric cell (a thin-walled vacuum-tight copper cell where the liquid sample is previously inserted and carefully weighed) was put on the sapphire disk and attached with a tiny amount of Apiezon vacuum grease. In this case, a calibrated carbon ceramic sensor thermometer (CCS A2) and a 1 kΩ chip as heating element were also attached to the calorimetric substrate.

To correctly subtract the contribution of the addenda to the measured heat capacities, and to obtain the wanted specific heat, one emptied calorimetric cell was also measured. In the rest of experiments, the small differences in the copper cell mass and Apiezon mass were also taken into account and corrections made in each case.

2.2 Experimental results

There are published measurements of the specific heat of toluene in the crystal state above 11 K by Scott et al. [24], and more recently of toluene glass by Yamamuro et al. [25] above 5.6 K, using adiabatic calorimetry. Those data are included in Fig. 2. The melting point of the crystal is $T_m$ = 178 K, and the glass-transition temperature was found to be $T_g$ = 117 K. Nevertheless, toluene is a bad glass-former and crystallizes readily. To suppress crystallization, Yamamuro et al. [25] doped toluene with about

10% of benzene. Effects of doping were corrected for by assuming the additivity of the heat capacities of toluene and benzene.

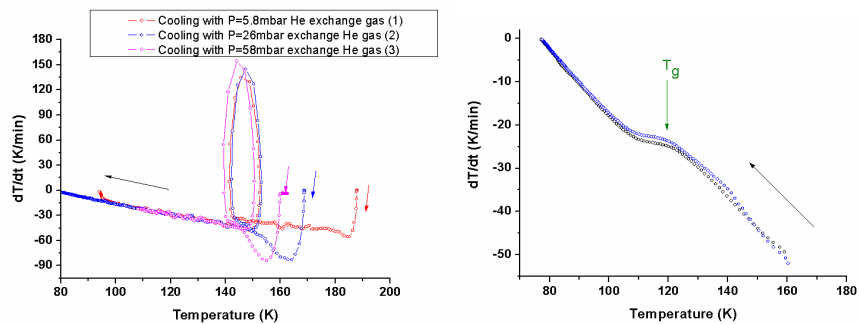

**Fig. 1** Thermograms of several fast cooling rates for the two cases studied. *Left*: Pure toluene, for three quenching attempts using different amounts of helium exhange gas and starting at different temperatures; in all cases crystallization unavoidably occurs at about 143 K. *Right*: Two different experiments by quenching the sample of 2% ethanol-doped toluene, using amounts of exchange gas similar to those used for pure toluene. No crystallization is observed, but a glass transition below 120 K is observed instead. (Color figure online)

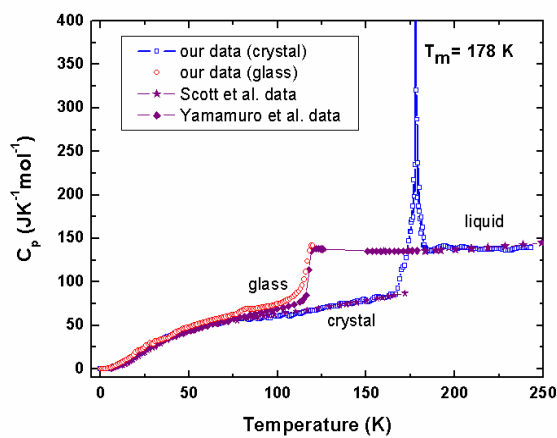

**Fig. 2** Specific heat in the whole measured range for the crystal (open squares) and glass (open circles) of toluene. Earlier published data for the crystal [24] (solid stars) and the glass [25] (solid lozenges) are also shown for comparison. (Color figure online)

In our first experiments, we filled-in the copper calorimetric cell with pure toluene. We measured the specific heat of the liquid and crystal states with liquid nitrogen as thermal bath, using the abovementioned continuous method. Then the liquid nitrogen bath was replaced by liquid helium and the curves were extended to lower temperatures (Fig. 2). Finally, more accurate single data points were measured using the thermal relaxation method (see Fig. 3) down to 4.2 K. A few additional points around 3 K were also obtained in a different experimental run. When using pure toluene, we only were able to obtain the crystal phase even quenching the liquid as fast as we could in our system. In Fig. 1 (*Left*) the time derivative of the temperature (cooling rate, going from the rightside to the leftside) is plotted. When cooling fast the liquid by suddenly introducing helium exchange gas in the internal vacuum chamber, trying different amounts of gas and starting at different temperatures, we achieved different cooling rates faster than −60 K/min. However, the liquid always irreversibly crystallized, as the big exothermic loops in the figure clearly demonstrate.

Therefore, in order to suppress crystallization, we prepared samples of doped toluene. But instead of putting 10% mol of benzene as Yamamuro et al. [25], we tried with ethanol but with only 2% molar fraction. The idea was that ethanol is a molecule with a shape more different to toluene than benzene, and hence less doping may be enough, thus minimizing the possible approximation errors involved in not using the pure substance. Furthermore, specific-heat data of ethanol are available [26-28] in a wide temperature range down to very low temperatures for a correct subtraction of its contribution. As shown in Fig. 1 (*Right*), we were able to cool it and obtain the glass phase avoiding the crystallization, passing through the glass transition at a moderate cooling rate of −25 K/min.

We measured the specific heat of a glass of toluene, obtained by slowly cooling of the supercooled liquid from a few K above $T_g$, and following the same procedures as with the crystal. Only, since we were especially interested in the low-temperature behavior of the glass state, we also measured carefully and more exhaustively the heat capacity down to 1.8 K, by pumping the liquid helium bath.

As one can see in Fig. 2, the agreement of our measurements in a wide temperature range with those earlier published data is very good. Further, in contrast to earlier adiabatic measurements, our continuous method allows to monitor and measure the phase transitions.

In Fig. 3 the raw heat capacity data at low temperatures (using the thermal relaxation method) is shown in a log-log plot, for the glass (red circles) and crystal (blue squares) phases. Solid symbols are the total measured heat capacities, the corresponding solid lines are the measured contribution of the addendas, and the empty symbols are the net values of toluene after subtraction. One can see that the contribution of the sample is significantly higher than that of the addenda (which is however not negligible at all), especially for the glass, as expected, and hence the sensitivity of the method is very reasonable.

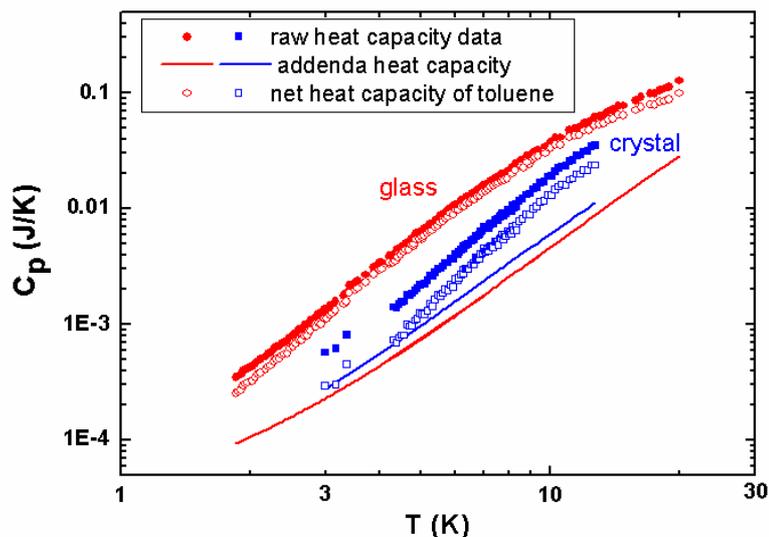

**Fig. 3** Raw heat-capacity data (solid symbols) measured at low temperatures with the thermal relaxation method. The measured heat-capacity curves of the corresponding addenda (including corrections for small differences in mass) are shown by solid lines. After subtraction of the addendas, net values of the heat capacity of the sample are obtained (open symbols). Data for the glass correspond to (red) circles, data for the crystal correspond to (blue) squares. (Color figure online)

In Fig. 4 (crystal) and Fig. 5 (glass), we present the obtained specific-heat data at low temperatures, in a typical $C_p/T$ vs $T^2$ plot, so that the Debye coefficients are given by their slope at the lower temperatures. This representation is also useful to perform a SPM fit for the glass data, as described in the next section. In both figures, earlier published data for the crystal [24] and for the glass [25] are also depicted.

Finally, we plot in Fig. 6 the Debye-reduced $C_p/T^3$ for both glass and crystal. Although very sensitive to data scatter, this is an useful representation to compare glasses and crystals since allows a direct visualization of glassy (TLS and boson peak) versus crystalline (flat Debye) behavior. As can be observed, the specific heat of the glass phase is much higher than that of the crystal. Toluene glass exhibits indeed a boson peak at 4.5 K, whereas the crystal shows an approximately flat behavior at lower temperatures with a shallow maximum at about 10 K, typical of molecular crystals [27-29].

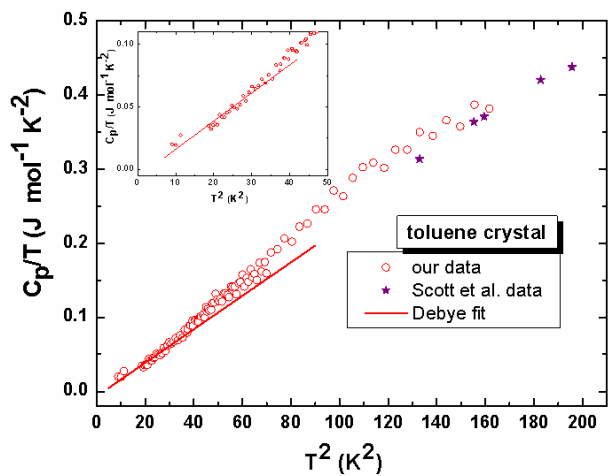

**Fig. 4** Specific heat of crystalline toluene (open circles) at low temperatures. Stars indicate earlier published data [24]. A least-squares linear Debye fit of the data at the lowest temperatures (see the inset) is shown by the solid line. (Color figure online)

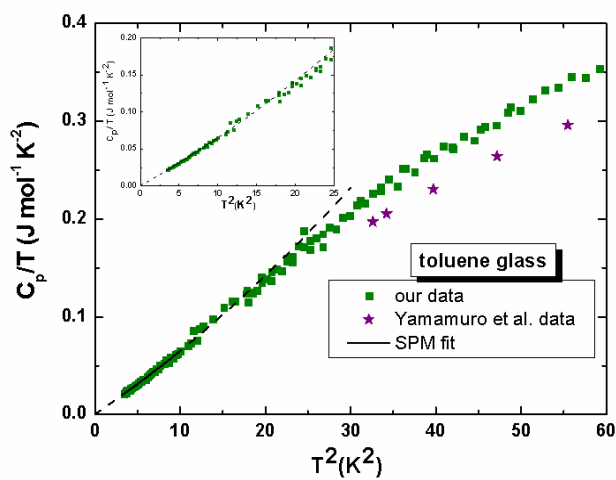

**Fig. 5** Specific heat of glassy toluene (solid squares) at low temperatures. Stars indicate earlier published data of benzene-doped toluene [25]. A least-squares quadratic fit of the data following the SPM at low enough temperatures (see the inset) is shown by the lines. (Color figure online)

## 3 Discussion

By means of the $C_p/T$ vs $T^2$ representation of our data at lower temperatures, we have a performed a Debye analysis for the crystal and a SPM one for the glass. In the former case (Fig. 4), a simple linear fit by least squares provides the Debye coefficient for the crystal $C_D = (2.26 \pm 0.09)$ mJ·mol$^{-1}$·K$^{-4}$, and hence a Debye temperature of $\Theta_D = 95.1$ K. To determine these *molecular* Debye temperatures, we have to consider in the Debye formula the number density of molecules rather than that of atoms, as previously discussed [30].

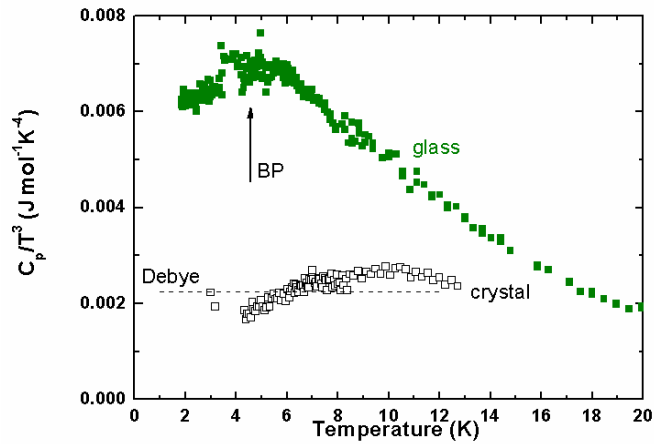

**Fig. 6** Debye-reduced $C_p/T^3$ data for both glass (solid symbols) and crystal (open symbols) of toluene. The boson peak (BP) at around 4.5 K for the glass and the estimated Debye level (averaged below 8 K) for the crystal are indicated (Color figure online)

For the glass, as mentioned above, we will employ a practical simplified version of the SPM [12,16] in order to analyze the data and quantify the low-temperature glassy anomalies. Below the boson peak maximum in $C_p/T^3$, it was shown that one can fit the data to

$$C_p = C_{TLS}T + C_D T^3 + C_{sm}T^5 \qquad (1)$$

where $C_{TLS}$ is the linear coefficient attributed to the tunneling TLS, $C_D$ is again the Debye coefficient due to lattice vibrations also present in glasses, and $C_{sm}$ is the contribution of the *soft modes* below the boson peak. Therefore a quadratic fit below $T = 3.5$ K in the $C_p/T$ vs $T^2$ representation directly provides the three coefficients. We have obtained (see the fit in Fig. 5 by the solid line, dashed lines indicating extrapolations): $C_{TLS}=0.6\pm1.8$ mJ·mol$^{-1}$·K$^{-2}$, $C_D=5.84\pm0.60$ mJ·mol$^{-1}$·K$^{-4}$, and $C_{sm}=0.062\pm0.045$ mJ·mol$^{-1}$·K$^{-6}$.

The first unexpected finding is the absence, within experimental error, of a linear term in the specific heat, ascribed to TLS. Nonetheless, specific-heat data at even lower temperatures or another physical property related to the presence of TLS (thermal conductivity, acoustic attenuation…) would be needed before concluding that toluene glass is indeed an exception to the universal behavior of glasses at very low temperatures, accounted for by the TM, as has been suggested [20,21].

The cubic coefficient provides us with the Debye coefficient (much higher than the corresponding crystal) and with corresponding Debye temperature for the glass, $\Theta_D$ = 69.3 K. This value is however very different from their claimed value of $\Theta_D$ =100.7 K by Yamamuro et al. [25]. Nevertheless, they did not obtain the value from $C_p$ data in the low-temperature limit but only after making some assumptions and fits assuming several rotational and vibrational contributions to the specific heat in the 5−89 K range. In our opinion, our method is much more direct and reliable.

In contrast to the apparent absence of TLS, toluene glass does exhibit a well-developed boson peak centered at around 4.5 K, as better observed in Fig. 6. For the many cases where both low-frequency vibrational spectroscopy and low-temperature specific heat data are available in the literature, a close relation between the boson peak position for $g(\nu)/\nu^2$ ($E_{BP}$ =$h\nu_{BP}$) and the maximum in $C_p/T^3$ at $T_{BP}$ is found, what was to be expected since the specific heat $C_v$ for non-metallic solids is directly related to the VDOS. To be more precise, one finds a numerical factor around 4−5 between both peaks, i.e. $E_{BP}$ = (4−5) $k_B T_{BP}$ [31,32]. Hence, from $T_{BP} \approx$ 4.5 K for glassy toluene one obtains $E_{BP} \approx$ (1.55−1.9) meV, in good agreement with the boson peak values reported by nuclear inelastic scattering (1.5 meV) [33] or by inelastic neutron scattering (1.5−3 meV) [34].

## 4 Summary and conclusions

In summary, we have measured the specific heat of toluene, both glass and crystal, in a wide temperature range from 1.8 K up to 250 K, thereby extending earlier published data to the low-temperature region where glassy anomalies appear.

With such an aim, we have found out that doping toluene with 2% mol of ethanol is enough to prepare a bulk sample in the glass state.

We have observed the expected Debye behaviour for the crystal and a clearly pronounced boson peak (at ca. 4.5 K) for the glass. However, the fitted linear coefficient of the specific heat was zero within our experimental error, pointing out an absence of the ubiquituous TLS in glasses. However, data at even lower temperatures, or a combination with other appropriate low-temperature measurements, should be performed in order to confirm or not the absence of TLS in amorphous toluene.

**Acknowledgements** This work has been partially supported by the Spanish Ministry of Economy through projects FIS2014-54498-R and MAT2014-57866-REDT, and through the "María de Maeztu" Programme for Units of Excellence in R&D (MDM-2014-0377), as well as by the Autonomous Community of Madrid through programme NANOFRONTMAG-CM (S2013/MIT-2850).

One of us (M.A.R.) is grateful to Anthony Leggett for discussions and for his suggestion to measure the specific heat of toluene.

**References**

1. P.W. Anderson, Science **267**, 1615 (1995)
2. R.C. Zeller, R.O. Pohl, Phys. Rev. B **4**, 2029 (1971)
3. W.A. Phillips (ed.), *Amorphous Solids: Low Temperature Properties* (Springer, 1981)
4. W.A. Phillips, J. Low Temp. Phys. **7**, 351 (1972)
5. P.W. Anderson, B.I. Halperin, C.M. Varma, Philos. Mag. **25**, 1 (1972)
6. U. Buchenau, N. Nücker, A.J. Dianoux, Phys. Rev. Lett. **53**, 2316 (1984)
7. U. Buchenau, M. Prager, N. Nücker, A.J. Dianoux, N. Ahmad, W.A. Phillips, Phys. Rev. B **34**, 5665 (1986)
8. V.K. Malinovsky, V.N. Novikov, P.P. Parshin, A.P. Sokolov, M.G. Zemlyanov, Europhys. Lett. **11**, 43 (1990)
9. M.A. Ramos, Phys. Rev. B **49**, 702 (1994)
10. U. Buchenau, Yu.M. Galperin, V.L. Gurevich, D.A. Parshin, M.A. Ramos, H.R. Schober, Phys. Rev. B **46**, 2798 (1992)
11. For a review, see D.A. Parshin, Phys. Rev. B **49**, 9400 (1994)
12. M.A. Ramos, U. Buchenau, Phys. Rev. B **55**, 5749 (1997)
13. L. Gil, M.A. Ramos, A. Bringer, U. Buchenau, Phys. Rev. Lett. **70**, 182 (1993)
14. V.L. Gurevich, D.A. Parshin, H.R. Schober, Phys. Rev. B **67**, 094203 (2003)
15. W. Schirmacher, Europhys. Lett. **73**, (2006) 892
16. M.A. Ramos, Phil. Mag. **84**, 1313 (2004)
17. C.C. Yu, A.J. Leggett, Comments Cond. Mat. Phys. **14**, 231 (1988)
18. A.L. Burin, D. Natelson, D.D. Osheroff, Y. Kagan, in *Tunnelling Systems in Amorphous and Crystalline Solids*, P. Esquinazi (ed.), Springer, Berlin, Chap. 3 (1998)

19. T. Pérez-Castañeda, C. Rodríguez-Tinoco, J. Rodríguez-Viejo, M.A. Ramos, PNAS **111**, 11275 (2014)
20. A.J. Leggett, D.C. Vural, J. Phys. Chem. B **117**, 12966 (2013)
21. I.Y. Eremchev, Y.G. Vainer, A.V. Naumov, L. Kador, Phys. Chem. Chem. Phys. **13**, 1843 (2011)
22. E. Pérez-Enciso, M.A. Ramos, Termochimica Acta **461**, 50 (2007)
23. T. Pérez-Castañeda, J. Azpeitia, J. Hanko, A. Fente, H. Suderow, M.A. Ramos, J. Low Temp. Phys. **173**, 4 (2013)
24. D.W. Scott et al., J. Phys. Chem. **66**, 911 (1962)
25. O.Yamamuro, I. Tsukushi, A. Lindqvist, S. Takahara, M. Ishikawa, T. Matsuo, J. Phys. Chem. B **102**, 1605 (1998).
26. O. Haida, H. Suga, S. Seki, J. Chem. Thermodyn. **9,** 1133 (1977)
27. C. Talón, M.A. Ramos, S. Vieira, Phys. Rev. B **66**, 012201 (2002)
28. M.A. Ramos, C. Talón, R. Jiménez-Riobóo, S. Vieira, J. Phys.: Condens. Matter **15**, S1007 (2003)
29. M. Hassaine, M.A. Ramos, A.I. Krivchikov, I.V. Sharapova, O.A. Korolyuk, R.J. Jiménez-Riobóo, Phys. Rev. B **85**, 104206 (2012)
30. A.I. Krivchikov, M. Hassaine, I.V. Sharapova, O.A. Korolyuk, R.J. Jiménez-Riobóo, M.A. Ramos, J. Non-Cryst. Solids **357**, 524 (2011)
31. A. I. Chumakov, A. Bosak, R. Rüffer, Phys. Rev. B **80,** 094303 (2009)
32. G. Carini, Jr. *et al*., Philosophical Magazine **96**, 761 (2016)
33. A.I. Chumakov et al., Phys. Rev. Lett. **92**, 245508 (2004)
34. I. Tsukushi et al., Journal of Physics and Chemistry of Solids **60**, 1541 (1999)